\documentclass[12pt]{amsart}
\usepackage{amssymb}
\usepackage{amsmath,amsthm,amsfonts}
\usepackage[applemac]{inputenc}
\usepackage{graphicx}
\usepackage{bbm}
\usepackage{pdfsync}
\usepackage{fancyhdr}
\usepackage{mathrsfs}
\usepackage{slashed}
\usepackage{color}
\usepackage{xcolor}
\usepackage[breaklinks,colorlinks,backref]{hyperref}

\hypersetup{
colorlinks, 
linktoc=all, 
linkcolor=blue, 
citecolor=hyptxt,
urlcolor=blue}
\hypersetup{
citebordercolor=,
filebordercolor=green,
linkbordercolor=blue
}
\definecolor{hyptxt}{rgb}{0.7, 0.4, 0.9}

\usepackage{bibtopic}

\usepackage[full]{textcomp} 
\usepackage{fbb} 
\usepackage[varqu,varl]{inconsolata}
\usepackage[libertine,bigdelims,vvarbb]{newtxmath}
\usepackage[cal=boondoxo]{mathalfa}
\usepackage[T1]{fontenc}
\useosf

\newtheorem{prop}{Proposition}[section]

\newcommand{\beprop}{\begin{prop}}
\newcommand{\enprop}{\end{prop}}
\newcommand{\bprf}{\begin{proof}}
\newcommand{\eprf}{\end{proof}}

\newcommand{\ket}[1]{|\kern.3ex#1\kern.3ex\rangle}
\newcommand{\bra}[1]{\langle\kern.3ex #1 \kern.3ex|}
\newcommand{\scalar}[2]{\langle\kern.3ex #1 \kern.3ex|\kern.3ex#2\kern.3ex\rangle}

\def\R{\mathbb{R}}

\def\C{\mathbb{C}}

\def\ii{\mathrm{i}}
\def\ud{\mathrm{d}}

\def\bu{\mathbbm{1}}

\def\ud{\mathrm{d}}
\definecolor{hervecolor}{rgb}{0.8,0,0.7}

\numberwithin{equation}{section}

\def\R{{\rm I\hspace{-.15em}R}}

\def\1{\mbox{I\hspace{-.15em}1}}

\def\b{\begin{equation}}
\def\e{\end{equation}}

\begin{document}
\date{\today}
\title{Quantum de Sitter geometry}
\author{M.V. Takook 
}

\address{\emph{ APC, UMR 7164}\\
\emph{CNRS, Astroparticule et Cosmologie, Universit\'e Paris Cit\'e, } \\
\emph{F-75013 Paris, France}}

\email{ takook@apc.in2p3.fr}

{\abstract{Quantum de Sitter geometry is discussed using elementary field operator algebras in Krein space quantization from an observer-independent point of view, {\it i.e.} ambient space formalism. In quantum geometry, the conformal sector of the metric becomes a dynamical degree of freedom, which can be written in terms of a massless minimally coupled scalar field. The elementary fields necessary for the construction of quantum geometry are introduced and classified. A complete Krein-Fock space structure for elementary fields is presented using field operator algebras. We conclude that since quantum de Sitter geometry can be constructed by elementary fields operators, the geometry quantum state is immersed in the Krein-Fock space and evolves in it. The total number of accessible quantum states in the universe is chosen as a parameter of quantum state evolution, which has a relationship with the universe's entropy. Inspired by the Wheeler-DeWitt constraint equation in cosmology, the evolution equation of the geometry quantum state is formulated in terms of the Lagrangian density of interaction fields in ambient space formalism.}}

\maketitle


\tableofcontents


\section{Introduction}

The quantum de Sitter geometry or quantum gravity is a subject fraught with enigmas that has garnered attention for over four decades. These enigmas encompass the absence of an S-matrix, challenges in defining observer-independent gauge-invariant, the issue of infrared divergences and renormalizability, and the construction of a complete space of quantum states, among others. 
In a previous article, we delved into the study of asymptotic states and the S-matrix operator, based on the construction of a complete Hilbert-Fock space for massive scalar fields in the de Sitter ambient space formalism \cite{tagahu}. The formulation of an observer-independent non-abelian gauge theory is also achievable using the ambient space formalism \cite{taga,ta231}. Krein space quantization leads to the disappearance of infrared divergence \cite{gareta00,ta22}. In this work, we explore the construction of a complete space of quantum states for quantum geometry and delve into quantum state evolutions.

Recently, Morris discussed that full quantum gravity may be perturbatively renormalizable in terms of Newton's constant, but non-perturbative in $\hbar$ \cite{morris}. Morris's interesting idea is to use the renormalization group properties of the conformal sector of gravity. It is well known that in quantum theory, the conformal sector of the spacetime metric becomes a dynamical degree of freedom as a result of the trace anomaly \cite{anmo,anmamo}. Then the metric-compatible condition is no longer valid and the simplest chosen geometry in this situation is Weyl or conformal geometry. Weyl geometry can be described with the tensor metric field $g_{\mu\nu}$ and its conformal sector, which can be expressed as a scalar field \cite{whe}.

In the Landau gauge of the gravitational field in de Sitter (dS) space, the conformal sector is described by a massless minimally coupled (mmc) scalar field $\phi_m$ \cite{ta09}. Its quantization with positive norm states breaks the dS invariant \cite{allen85}. For its covariant quantization, Krein space quantization is needed \cite{gareta00}. Using the interaction between the gluon field and the conformal sector of the metric in Krein space quantization, the axiomatic dS quantum Yang-Mills theory with color confinement and the mass gap can be constructed \cite{taga,ta231}. We showed that the mmc scalar field can be considered as a gauge potential and the dS metric field and its conformal sector are not elementary fields \`a la Wigner sense  \cite{ta223}. However, they can be written in terms of elementary fields, in which the mmc scalar field plays a central role. We presented two different perspectives on quantum geometry, namely the classical and quantum state perspectives. The first is observer-dependent and the second is observer-independent. We discussed that it is essential to use an observer-independent formalism when considering quantum geometry. Therefore, we must use the algebraic method in the ambient space formalism for studying quantum geometry and define the quantum state of geometry $|\mathcal{G}\rangle$, which will be addressed in this paper.

In recent years, some authors have also used the idea of the algebraic approach to consider quantum gravity. This approach takes into account an algebra of observables, Hilbert space structure, and geometry quantum state \cite{chlopewi,pewi}. By using the algebraic method, in the previous paper the complete Hilbert-Fock space was constructed for the massive elementary scalar field in dS ambient space formalism \cite{tagahu}. Here we generalize it to construct a Hilbert-Fock space structure for any spin fields in subsection \ref{HS}. This space is a complete space under the action of all elementary field operators in dS space except linear gravity and the mmc scalar field. To obtain a complete space for these two fields, we need Krein space quantization, which is discussed in subsection \ref{sbsekr}. We know that the QFT in Krein space quantization combined with light-cone fluctuation is renormalizable \cite{ta22}. Therefore, the two problems of renormalizability and constructing the complete space of quantum states for quantum dS geometry can be solved using Krein space quantization and ambient space formalism.

In the next section, we briefly review the necessary notions of general relativity and QFT for our discussion. All possible fundamental fields necessary for quantum geometry are introduced and classified in Section \ref{elementary}. In section \ref{qgss}, Krein-Fock space as a complete space for quantum geometry is presented. The quantum state of geometry $|\mathcal{G}\rangle$ is considered in section \ref{qse}, which can be formally written in terms of orthonormal bases of Krein-Fock space. It is immersed and evolves in the Krein space $\mathcal{K}$ instead of the Hilbert space $\mathcal{H}$. Quantum state evolution is characterized by the total number of accessible quantum states in the universe, which has a relationship with the total entropy of the universe. In Section \ref{qse}, using the Wheeler-DeWitt equation, the constraint equation for the quantum state of geometry is formulated in terms of the Lagrangian density of interaction fields.


\section{Basic notions} \label{notion}

Spacetime structure and observation are challenging concepts in quantum theory. Riemannian geometry is usually employed in general relativity. In Riemannian geometry, spacetime can be described by the metric $g_{\mu\nu}(\vec x,t)$ (with the metric-compatible condition) and curved spacetime can be visualized as a $4$-dimensional hypersurface immersed in a flat spacetime of dimensions greater than $4$. Although the $4$-dimensional classical spacetime hypersurface is unique and observer-independent, the choice of metric $g_{\mu\nu}$ is completely observer-dependent, which is a manifestation of the general relativity principle, {\it all observers are equivalent} ({\it i.e.} diffeomorphism covariance). However, spacetime hypersurfaces are no longer unique in quantum geometry. In the classical perspective, quantum spacetime is described by a sum of different spacetime hypersurfaces \cite{ta223,ta20}. But in the quantum perspective, the quantum spacetime is modelled by a quantum state $|\mathcal{G}\rangle$, which is presented in Section \ref{qgss}.

In QFT, the physical system can be described by a quantum state vector $|\Psi(\nu,n)\rangle$, where $\nu$ and $n$ are the set of continuous and discrete quantum numbers respectively. They are labeled the eigenvector of the set of commutative operator algebras of the physical system and determine the Hilbert space, for dS space with more details see \cite{tagahu}. Although the particle and tensor(-spinor) fields, $\Phi(\vec x,t)$, are immersed in a spacetime manifold $M$, the quantum state vector is immersed in a Hilbert space $ \mathcal{H}$. The field operator $\Phi(\vec x,t)$ plays a significant role in the connection between these two different spaces: a spacetime manifold $M$ and a Hilbert space $ \mathcal{H}$. On the one hand, it is immersed in spacetime, and on the other hand, it acts in Hilbert's space, which is defined at any point in a fixed classical spacetime background $M$ (of course in the distribution sense). Hilbert space can be thought of as the "fiber" of a bundle over the spacetime manifold, where each point of the manifold corresponds to a different fiber, $ \mathcal{H}\times M$. The bundle is typically referred to as a "Fock space bundle". For a better understanding of this idea, see \cite{gui} and noncommutative geometry \cite{chsu}. The Wightman two-point function, $\mathcal{W}(x,x')=\langle \Omega |\Phi(x)\Phi(x')|\Omega\rangle$, provides a correlation function between two different points in spacetime and their corresponding Hilbert spaces. $|\Omega\rangle$ is the vacuum state. Historically, time played a central role in quantum theory, since the time parameter describes the evolution of the quantum state. Time, however, is an observer-dependent quantity in special and general relativity, and for quantum geometry to be observer-independent, the time evolution of quantum states must be replaced by another concept, which is discussed in Section \ref{qse}.

It is useful to recall that in contrast to all massive and massless elementary fields, the mmc scalar field disappears at the null curvature limit \cite{ta14}. Its quantization with positive norm states also breaks the dS invariant \cite{allen85} and its behavior is very similar to the gauge fields \cite{ta223}. Since the collection of all these properties is the same as the properties of curved space-time geometric fields, the mmc scalar field can be considered as part of space-time geometry. This idea was previously applied to explain the confinement and mass gap problems in dS quantum Yang-Mills theory, by using the interaction between the vector field and the scalar gauge field, as a part of the spacetime gauge potential \cite{ta231,khletu}.


\section{Elementary fields} \label{elementary}

In the background field method, $ g_{\mu\nu}=g_{\mu\nu}^{bg}+h_{\mu\nu}$, the linear gravity $h_{\mu\nu}$ propagate on the fixed background $g_{\mu\nu}^{bg}$. The tensor field $h_{\mu\nu}$ can be divided into two parts: the traceless-divergencelessness part $h_{\mu\nu}^T$, which can be associated with an elementary massless spin-$2$ field and the pure trace part, $h_{\mu\nu}^P=\frac{1}{4}g_{\mu\nu}^{bg} \phi_m$. The pure trace part can be transferred to the conformal sector of the background metric:
\b \label{consec}  g_{\mu\nu}=g_{\mu\nu}^{bg}+h_{\mu\nu}^T+h_{\mu\nu}^P=\left(1+  \frac{1}{4} \phi_m\right)g_{\mu\nu}^{bg}+ h_{\mu\nu}^T \equiv e^{\sigma(x)} g_{\mu\nu}^{bg}+ h_{\mu\nu}^T\, .\e
It is also called the conformal sector of the metric, which becomes a dynamical variable in quantum theory \cite{anmo,anmamo}. Quantum geometry is equal to the quantization of the tensor field $g_{\mu\nu}$ or equivalently $g_{\mu\nu}^{bg}$, $h_{\mu\nu}^T$, $h_{\mu\nu}^P$ and $\phi_m$. In quantum geometry, the choice of the curved metric background $g_{\mu\nu}^{bg}$ is not critical since we have simultaneous fluctuations in $g_{\mu\nu}^{bg}$ and $h_{\mu\nu}$ and it can also be considered as an integral over all possible spacetime hypersurfaces \cite{ta223,ta20}. For a covariant quantization of $h_{\mu\nu}$, the background must be curved \cite{bida}, and from the cosmological experimental data, the dS metric $g_{\mu\nu}^{dS}$ is selected as a curved spacetime background. 

For an observer-independent point of view, we use the dS ambient space formalism \cite{taga,ta14}. In this formalism, the dS spacetime can be identified with the $4$-dimensional hyperboloid embedded in the $5$-dimensional Minkowski spacetime as:
\b \label{dSs} M_H=\left\{ x^\alpha\equiv x \in \R^5| \; \; x \cdot x=\eta_{\alpha\beta} x^\alpha
x^\beta =-H^{-2}\right\}\,,\;\; \alpha,\beta=0,1,2,3,4\,, \e
with $\eta_{\alpha\beta}=$diag$(1,-1,-1,-1,-1)$ and $H$ is like Hubble's constant parameter. The metric is:
\b \label{dsmet} \ud s^2=g_{\mu\nu}^{dS}\ud X^{\mu}\ud X^{\nu}=\left.\theta_{\alpha\beta}\ud x^{\alpha}\ud x^{\beta}\right|_{x\cdot x=-H^{-2}}\,,\e
where the $X^\mu$'s ($\mu=0,1,2,3$) form a set of $4$-space-time intrinsic coordinates on the dS hyperboloid, and the $x^{\alpha}$'s are the ambient space coordinates. In this coordinate, the transverse projector on the dS hyperboloid, $ \theta_{\alpha \beta}=\eta_{\alpha \beta}+ H^2x_{\alpha}x_{\beta}$, plays the same role as the dS metric $g_{\mu\nu}^{dS}$. In this formalism, quantum geometry is described by the quantization of the tensor fields $\theta_{\alpha\beta}, \mathcal{K}_{\alpha\beta}^T$ and  $\mathcal{K}_{\alpha\beta}^P=\frac{1}{4}\theta_{\alpha\beta}\phi_m $ ($x\cdot \mathcal{K}^T=0=x\cdot \mathcal{K}^P $). 

Although the tensor field $\mathcal{K}_{\alpha\beta}^T$ and scalar field $\phi_m$  are elementary fields, the background metric $\theta_{\alpha\beta}$ and conformal sector of the metric, $\mathcal{K}_{\alpha\beta}^P=\frac{1}{4}\phi_m \theta_{\alpha\beta}$, are not elementary fields  \`a la Wigner sense, since \cite{ta223}:
\b \label{bmcc} \theta_\alpha^\alpha=4, \;\;  \mathcal{K}_\alpha^{P\alpha}=\phi_m\,, \;\; \nabla^\top\cdot \mathcal{K}_{\beta}^P=\frac{1}{4}\partial^\top_\beta \phi_m \, .\e
The transverse-covariant derivative acting on a tensor field of rank-$2$ is defined by:
\begin{equation}\label{dscdrt}
\nabla^\top_\alpha K_{\beta\gamma}\equiv \partial^\top_\alpha
K_{\beta\gamma}-H^2\left(x_{\beta}K_{\alpha\gamma}+x_{\gamma}K_{\beta\alpha}\right)\,,
\end{equation} 
where $\partial_\beta^\top =\theta_{\alpha \beta}\partial^{\alpha}=
\partial_\beta + H^2 x_\beta x\cdot\partial$ is tangential derivative. The tensor fields $\mathcal{K}_{\alpha\beta}^P$ and $\theta_{\alpha \beta}$ can be written in terms of elementary fields: the massive rank-$2$ symmetric tensor field $K_{\alpha\beta}^\nu$ ($\nu^2=\frac{15}{4}$), mmc scalar gauge field $\phi_m$ and massless vector field $A_\alpha=\partial^\top_\alpha \phi_m$ \cite{ta223}. 

The tensor field $K_{\alpha\beta}^\nu$ is discussed as massive gravity in literature, which was studied in the previous paper \cite{gagata}. The massless vector field quantization was presented in \cite{gagarota}. The constant pure trace part evokes the famous zero-mode problem in linear quantum gravity and the quantization of the mmc scalar field. The classical structure of our universe may be constructed by the following fundamental fields, which can be divided into three categories:
\begin{itemize}
\item{A: Massive elementary fields with different spins, which transform by the unitary irreducible representation (UIR) of the principal series of the dS group.}
\item{B: Massless elementary fields with the spin $s\leq 2$, which includes the gravitational waves $\mathcal{K}_{\alpha\beta}^T$, and mmc scalar fields $\phi_m$. They transform by the indecomposable representation of the dS group where the central part is the discrete series representation of the dS group. They play an important role in defining the interaction between different fields as the gauge potential \cite{ta223}.}
\item{C: The spacetime geometrical fields $\theta_{\alpha\beta}$ and conformal sector of the metric $\mathcal{K}_{\alpha\beta}^P$, which are not elementary fields, but they can be written in terms of the elementary fields of categories A and B. Although in classical field theory, they preserve the dS invariant, their quantization breaks the dS spacetime symmetry \cite{ta223}. }
\end{itemize}
The quantization of the elementary massive and massless fields with the spin $s\leq 2$ in dS ambient space formalism has been previously constructed for principle, complementary, and discrete series representations of the dS group; for a review, see \cite{ta14}. The mmc scalar field and linear gravity $\mathcal{K}_{\alpha\beta}^T$ can be quantized in a covariant way in Krein space quantization \cite{ta09,gareta00}. We know that the QFT in curved spacetime suffers from renormalizability, and for solving this problem, Krein space quantization must be used; see \cite{ta22} and the references therein. Due to the quantum fluctuation of tensor field $\mathcal{K}^P_{\alpha\beta}\,$, the dS invariant is broken \cite{ta223}, which is reminiscent of the quantum instability of dS spacetime \cite{anilto}.


\section{Quantum space of states} \label{KFS}

Before discussing the quantum geometry space of states, Hilbert-Fock space and Krein-Fock space constructions are briefly introduced using dS group algebra and field operators algebra. We discuss that the Krein-Fock space is a complete space for all elementary fields and quantum geometry.

\subsection{Hilbert-Fock space} \label{HS}

Fro the dS spacetime, one can construct a one-particle Hilbert space $\mathcal{H}^{(1)}$ from dS group algebra for the principal, complementary, and discrete series UIR of the dS group \cite{tho,dix,tak}:
\b \label{algebra} [J_a,J_b]=f_{ab}^c J_c \;\; \Longrightarrow \;\; |\nu; j_1,j_2; m_{j_1},m_{j_2}\rangle \in \mathcal{H}^{(1)}\equiv \bigoplus_{\nu; j_1 ,j_2} \; \mathcal{H}^{\nu\,; j_1,j_2}\, ,\e
where $J_a$  are the generators of the de~Sitter group ($a,b=1,2,\cdots,10$), $f_{ab}^c$ are the structure constants,  $j_1$ and $j_2$ are two numbers, labeling the UIR's of the maximal compact subgroup SO($4$), picked in the sequence $0,\frac{1}{2},1,\cdots$, such that $-j_1\leq m_{j_1} \leq j_1$. The $\nu$'s are sets of parameters numbering the columns and rows of the (generalized) matrices, assuming continuous or discrete values \cite{tho}. 

The UIRs of the principal and complementary series are classified by the two parameters $j$ and $\nu$, whereas $\nu$ is continuous and the sum is replaced with an integral \cite{tagahu,tak,moy}:
\b \label{pral} \mathcal{H}^{(1)}\equiv \bigoplus_{ j}  \int_0^\infty \ud \nu\; \rho(\nu)\; \mathcal{H}^{\nu\,;j}\equiv \bigoplus_{ j} \mathcal{H}^j \, ,\e
where $\varrho(\nu)$ is a positive weight in the dS background \cite{brmo96}. $\nu$ refers to the mass parameter and $j$ is equivalent to the spin. They determine the eigenvalues of the Casimir operators of the dS group. $ \mathcal{H}^j\equiv \int_0^\infty \ud \nu\; \rho(\nu)\; \mathcal{H}^{\nu\,;j}$ is one-particle Hilbert space for a specific spin $j$. A quantum state in this Hilbert space may be represented with $|\nu,j \,; L\rangle  \in \mathcal{H}^{(1)}$, where $L$ is a set of quantum numbers that concern the maximal set of commuting operators with the Casimir operators, which represent the dS enveloping algebra \cite{tagahu,moy}. It is critical to note that the Hilbert space $\mathcal{H}^{\nu\,;j}$ is not a complete space under the action of the dS group generator $J_a$, but the Hilbert space $\mathcal{H}^{(1)}$ is complete space \cite{tagahu}. The notation $\mathcal{H}^{(1)}$ means it is the ``one particle state'' (first quantization). Since dS group generators and field operators do not modify the spin of the state, for a fixed spin $j$, the Hilbert space $ \mathcal{H}^j$ is also a complete space. In this study, we do not consider supersymmetry and supergravity, otherwise, the sum over the index $j$ should be taken into account for obtaining a complete space.

There are different realizations for the bases of the one-particle Hilbert space $\mathcal{H}^{(1)}$ where some of them are presented for the scalar field ($j=0$) in \cite{tagahu}. Formally, we define a UIR of the de Sitter group by $U^{(\nu;j)}(g(\alpha_a)) \equiv U(J_b\,, \alpha_a, \nu,j)$, which is a regular function of dS group generators and acts on the Hilbert space as:
\b \label{geact} |\psi\rangle \in  \mathcal{H}^{(1)} \Longrightarrow  U(J_b\,, \alpha_a, \nu,j) |\psi\rangle= |\psi' ;\alpha_a,\nu,j\rangle \in \mathcal{H}^{(1)}\, , \e
where the $10$ $\alpha_a$'s are the group parameters. These parameters make up a $10$-dimensional topological space $\mathcal{T}(\alpha_a)$. By using the expressions of the matrix elements ($\sim$ coefficients) of this representation,
\b \label{clasfield} |\psi\rangle\, , |\psi'\rangle \in  \mathcal{H}^{(1)} \,, \;\;\; \langle \psi | U^{(\nu;j)}(g(\alpha_a)) |\psi'\rangle\, ,\e
one can construct a space of square-integrable functions over some subspaces $ \mathcal{S}$ of the topological space $\mathcal{T}$, {\it i.e.} $L^2(\mathcal{S})$ where $ \mathcal{S} \subset\mathcal{T}(\alpha_a)$. Takahashi discusses different subspaces and defines the relations between some of them by the Plancherel formula \cite{tak}. The classical tensor(-spinor) field can be identified with some coefficients of the UIR of the dS group in dS ambient space coordinates under certain conditions: $\Phi(x) \approx   \langle \psi | U^{(\nu;j)}(g(\alpha_a)) |\psi'\rangle$, where $ x\in M_H$. 

In QFT, these classical fields are assumed to be the operators, which act on a space with the Fock structure, {\it i.e.} like the harmonic oscillators. The well-defined operators are defined in a tempered-distributional sense on an open subset $\mathcal{O}$ of spacetime \cite{brmo96}:
\b \label{algebcom2} \Phi(f)=\int \ud\mu(x) \,f(x) \Phi(x) \, ,\e
where $f$ is a test function and $\ud\mu(x)$ is  dS invariant measure element. As usual in Fock structure, the field operator can be written in terms of its creation part, $\Phi^+$, and its annihilation parts $\Phi^-$: $ \Phi(f)=\Phi^-(f)+\Phi^+(f)$, where $\Phi^+(f)$ creates a state and $\Phi^-(f)$ annihilates a state in the Fock space. By defining a ``number'' operator $N(f,g)\equiv \Phi^+(f)\Phi^-(g)$, one can prove the following algebra, which results in the construction of the Hilbert-Fock space \cite{tagahu}:
\b \label{gcsll} \left\{\begin{array}{clcr} \left[\Phi^-(f), \Phi^+ (g)\right] &= \mathcal{W}(f,g) \bu \,,\\
\left[N(f,g), \Phi^+ (k)\right]&=\mathcal{W}(g,k) \Phi^+(f) \, ,\\
\left[N(f,g), \Phi^-(k)\right]&= -\mathcal{W}(k,f) \Phi^-(g) \, ,
\end{array} \right.
\e
where $\mathcal{W}(f,g)=\int \ud\mu(x)\ud\mu(x')f(x)\mathcal{W}(x,x')g(x')$, and here $\Phi$ is the tensor field. For the fermion field, the anti-commutation relation must be used. $\mathcal{W}(x,x')=\langle \Omega |\Phi(x)\Phi(x')|\Omega\rangle$ is the Wightman two-point function and $|\Omega\rangle$ is the vacuum state, which can be fixed in the null curvature limit \cite{brmo96}. 

Now using the infinite-dimensional closed local algebra \eqref{gcsll}, one can construct the Hilbert-Fock space in a distributional sense on an open subset $\mathcal{O}$ of the dS spacetime \cite{gui,brmo96}:
\begin{equation} \label{hfss} \mathcal{H}\equiv \mathcal{F}(\mathcal{H}^{(1)})=\left\{ \C, \mathcal{H}^{(1)}, \mathcal{H}^{(2)},\cdots, \mathcal{H}^{(n)}, \cdots \right\} \equiv \bigoplus_0^n \mathcal{H}^{(n)}\equiv e^{\mathcal{H}^{(1)}} \,, \end{equation}
where $\C$ is vacuum state, $\mathcal{H}^{(1)}$ is one-particle states and $\mathcal{H}^{(n)}$ is n-particle states. The n-particle states are constructed by the tensor product of one-particle states (for bosons, a symmetry product, $ \mathcal{H}^{(2)}=S\mathcal{H}^{(1)}\otimes \mathcal{H}^{(1)}$ and for fermions anti-symmetric products, $ \mathcal{H}^{(2)}=A\mathcal{H}^{(1)}\otimes \mathcal{H}^{(1)}$). We used the Hilbert-Fock space name to emphasize that the structure of our QFT Hilbert space is in the form of the equation \eqref{hfss}. An overview of axiomatic quantum fields, observable algebraic nets, and the algebraic setting of second quantization can be found in \cite{gui}. Considering the interaction fields, it does not add any supplementary operators but reduces the number of commuting operators. Then we have a new algebra, resulting in a new Hilbert space $\mathcal{H}_{int}$. This space $\mathcal{H}_{int}$ can be immersed in the original space, which means $\mathcal{H}_{int}\subset \mathcal{H}$. Therefore one can use the Hilbert space \eqref{hfss} for the interaction fields, for the scalar field see \cite{tagahu}.


\subsection{Krein-Fock space} \label{sbsekr}

The above Hilbert-Fock space structure cannot be used for the mmc scalar field operator, and then for $\mathcal{K}_{\alpha\beta}^T$, $\theta_{\alpha\beta}$, and $\mathcal{K}_{\alpha\beta}^P$. The one-particle Hilbert space of mmc scalar field is not a complete space under the action of the dS group generators $J_a$. Their action results in the negative norm state \cite{gareta00}. This problem appeared as a dS-invariant breaking and the appearance of infrared divergence in the two-point function $\mathcal{W}(x,x')$ \cite{allen85}. Then the field operator algebra \eqref{gcsll} breaks the dS invariant and the dS invariant Hilbert-Fock space structure cannot be constructed. That means the effect of the field operator over some states results in states out of the Hilbert space, {\it i.e.} states with the negative norm. These states are necessary to obtain a complete space.

This problem was solved in Krein space quantization, which is a direct tensorial sum of a Hilbert space and its anti-Hilbert space \cite{gareta00}:
\b \label{onepks} \mathcal{K} ^{(1)}\equiv \mathcal{H}^{(1)} \oplus [\mathcal{H}^{(1)}]^*\,.\e
In this case, the two-point function is the imaginary part of the two-point function of the positive mode solutions \cite{ta3,ta02}:
\b \label{ktpf} \mathcal{W}_k(x,x')=\mathcal{W}(x,x')+\mathcal{W}_n(x,x')= \ii \mathrm{Im} \mathcal{W}(x,x')\,,\e
which is dS invariant. $\mathcal{W}_n(x,x')=-\mathcal{W}^*(x,x')$ is the two-point function of the negative norm states. If we replace the two-point function in the field operator algebra \eqref{gcsll} with the Krein two-point function $\mathcal{W}_k(x,x')$, we can construct the following dS invariant Krein-Fock space structure:
\b \label{fockr} \mathcal{F}(\mathcal{K}^{(1)})=\left\{ \C, \mathcal{K}^{(1)}, \mathcal{K}^{(2)},\cdots, \mathcal{K}^{(n)}, \cdots \right\}  \equiv \bigoplus_{n=0}^\infty \mathcal{K}^{(n)}\equiv e^{\mathcal{K}^{(1)}}\,.\e
It is pertinent to note that the Krein-Fock space is a complete space for all massive and massless elementary field operators in the dS spacetime. The Krein space can be considered the "fiber" of a bundle over the dS base manifold, $ \mathcal{K}\times X_H$. In this complete space, we can define (in the distribution sense) the identity operator formally as $ \bu\equiv \sum_\mathcal{M}| \mathcal{M}\rangle \langle \mathcal{M}| $.


\subsection{Quantum geometry space of states} \label{qgss}

In quantum geometry, the biggest challenge appears in the quantization of $\theta_{\alpha\beta}$. Its quantum fluctuation breaks the dS invariant and the concept of spacelike separation points cannot be defined. Therefore one cannot define the field operator algebra \eqref{gcsll} for the construction of the Krein-Fock space structure. This problem has a long history and we do not want to discuss it here, see \cite{ta20}. We ignore this problem for now since the Krein-Fock space \eqref{fockr} is a complete space for all elementary fields in dS space, and the geometrical fields $\theta_{\alpha\beta}$ and $\mathcal{K}^P_{\alpha\beta}$ can be written in terms of elementary fields. Therefore we can use the Krein-Fock space \eqref{fockr} for quantum field operators $\theta_{\alpha\beta}$ and $\mathcal{K}^P_{\alpha\beta}$. We can assume that quantum geometry is described by a quantum state $|\mathcal{G}\rangle$, which is immersed in the Krein-Fock space \eqref{fockr}, $|\mathcal{G}\rangle\in\mathcal{F}(\mathcal{K}^{(1)})$. It can be formally written by a superposition on the Krein-Fock space bases in the following form:
 \b  \label{qsog} |\mathcal{G}\rangle=\sum_\mathcal{M} c_\mathcal{M}(\mathcal{G})|\mathcal{M}\rangle\, .\e
 The action of the field operators $\theta_{\alpha\beta}$ on $ |\mathcal{G}\rangle$ results in a new state $|\mathcal{G}'\rangle$, which is in the Krein-Fock space: $|\mathcal{G}'\rangle=\theta_{\alpha\beta}|\mathcal{G}\rangle\in\mathcal{F}(\mathcal{K}^{(1)}) $.

Krein-Fock space in quantum geometry plays the same role as all parts of the dS spacetime hyperboloid in classical theory. Hilbert space $\mathcal{H}$ may be considered the observable part of space for an observer. When we discuss Hilbert space this means we have only positive norm states. First, let's review some facts about dS spacetime, in which particles and fields are immersed and evolve within. The basis vectors of dS spacetime have negative, positive and null norms, where the spacetime interval is given by the metric signature $(1,-1,-1,-1)$. When we move from Euclidean geometry to Minkowskian geometry, negative norm vectors appear. However, this norm's meaning is completely different from the Euclidean norm. There are three types of vectors in spacetime based on their norms: light-like vectors, space-like vectors, and time-like vectors. Some regions of the dS hyperboloid are also not observable to an observer due to spacetime curvature and the event horizon. However, these regions are necessary to construct a covariant formalism of the physical system.

Similarly, in discussions of quantum geometry, we must employ the quantum state with a negative norm for covariant quantization. Consequently, the Krein-Fock space constitutes a complete space under the influence of geometrical field operators. However, the physical significance of this negative norm state in quantum geometry remains an open question. In classical dS geometry, certain spacetime regions are beyond the observation of an observer. Analogously, in dS quantum geometry, negative norm states are necessary to achieve a complete space, yet they remain unobservable to a local observer. By implementing the observer reality principle \cite{ta22}, these states can be excluded from the observer's physical space. It can be argued that the absence of interaction beyond the event horizon parallels the lack of interaction between negative and positive norm states in Krein's space for a local observer. Similar to the negative values in the Wigner quasi-probability distribution function in quantum optics, which signify non-classical states and quantum interference effects, the negative norm states in the Krein quantization of geometry might represent the non-classical and pure quantum interference phenomena, devoid of a direct classical counterpart.

At the null curvature limit, negative norm states have negative energy \cite{ta02}. For a free particle state in flat spacetime, they have no physical interpretation and can be considered auxiliary states for the local observer. If we assume that the gravity state contains negative energy, the matter-radiation state carries positive energy, and their sum is zero, this hypothesis is compatible with the theory of the creation of everything from nothing in cosmology, see the similar discussion after equation \eqref{evoph}.

Different quantum gravity models are constructed in Hilbert space rather than Krein space. One of them, which is very close to our model is noncommutative geometry \cite{chsu}, where in the previous paper some similarities and differences were discussed \cite{ta223}. The other is higher-dimensional spacetime $M_d$ with $d>4$. In this case, the field operator algebra \eqref{gcsll} can be defined concerning the space-like separation point in $M_d$, which can be imagined as a fixed background space. The quantum fluctuation of $X_4$ may be considered as a sum over different $4$-dimensional manifolds in $M_d$. However higher-dimensional spacetime is used in some theoretical models.


\section{Quantum state evolution} \label{qse}

As time is an observer-dependent quantity, time evolution does not make sense in quantum geometry from an observer-independent point of view. We see that the Kerin-Fock space is constructed from the free field operators algebra, which explains the kinematics of the physical system.  Since all matter-radiation fields and geometrical fields are entangled and the change of one has a consequence for the other, therefore the dynamics of a physical system may be extracted from the algebra of interaction fields. But here for simplicity, we use the Lagrangian density of the interaction field for defining the evolution equation of the geometry quantum state.

Assuming the universe's evolution begins from the vacuum state, {\it i.e.} a quantum state without any quanta of the elementary and geometrical fields, $\vert \mathcal{G}_i\rangle\equiv \vert \Omega \rangle$. Our universe is also assumed to be an isolated system. By these assumptions, the universe began with zero entropy. Due to quantum vacuum fluctuations in all elementary fields, and the interaction between some of them in the creation situation, the universe leaves the vacuum state and enters the inflationary phase. This means its entropy increases because isolated systems spontaneously evolve toward thermodynamic equilibrium, which is a state of maximum entropy. In the inflationary phase, which is explained by dS spacetime, we have an infinite-dimensional Hilbert space. But due to the compact subgroup SO($4$) of the dS group and the uncertainty principle, the total number of quantum one-particle states becomes finite \cite{taen}. The finiteness hypothesis of energy results in the finiteness of the total number of quantum states $\mathcal{N}$ in Fock space, which results in a finite entropy for the universe \cite{taen}.

Since the universe is an isolated system and its entropy is increasing, $\mathcal{N}$ increases with the evolution of the universe. Therefore the total number of accessible quantum states in the universe, $\mathcal{N}$, may play the role of the time parameter and is used as the parameter of quantum state evolution. We assume that the evolution of the quantum state can be written by an operator $\mathsf{U}$ as follows:
\b \label{teod} |\mathcal{G}\,;\mathcal{N}\rangle\in\mathcal{F}(\mathcal{K}^{(1)})  \Longrightarrow  \mathsf{U}(\mathcal{N}',\mathcal{N})\vert \mathcal{G}\,;\mathcal{N}\rangle\equiv \vert \mathcal{G}'; \mathcal{N}'\rangle \in\mathcal{F}(\mathcal{K}^{(1)})\, ,\e
which satisfies the following conditions:
\b \label{evolu}  \mathsf{U}(\mathcal{N}_3,\mathcal{N}_2)\mathsf{U}(\mathcal{N}_2,\mathcal{N}_1)=\mathsf{U}(\mathcal{N}_3,\mathcal{N}_1)\,,\; \;\; \mathsf{U}(\mathcal{N},\mathcal{N})=\bu \,. \e
Due to the principle of increasing entropy, we always have $  \mathcal{N}_3\geq \mathcal{N}_2\geq \mathcal{N}_1$. For obtaining the evolution operator $U(\mathcal{N}',\mathcal{N})$, we need a constraint equation for the quantum state.

The quantum state of the universe is a function of the configuration of all the fundamental fields in the universe, Section  \ref{elementary}. Previously, we obtained these fields' classical action or Lagrangian density in the ambient space formalism. It can be formally written in the following form:
\b \label{unlade}  S[\Phi]=\int  \ud\mu(x) \mathcal{L}(\Phi,\nabla^\top_\alpha\Phi)=\int  \ud\mu(x) \left[\mathcal{L}_f(\Phi,\nabla^\top_\alpha\Phi)+\mathcal{L}_{int}(\Phi,\nabla^\top_\alpha\Phi)\right]\,.\e
For free field Lagrangian density $\mathcal{L}_f$ see \cite{ta14}, and for interaction case $\mathcal{L}_{int}$ see \cite{taga,ta14}. Since in dS spacetime $x^0$ plays the same role as the time variable in Minkowski space, see section $4$ in \cite{tagahu}, we define the conjugate field variable by $\Pi \equiv \nabla^\top_0\Phi$. The Legendre transformation of the Lagrangian density $\mathcal{L}(\Phi,\nabla^\top_\alpha\Phi)$ with respect to the variable $\nabla^\top_0\Phi$ can be rewritten in the following form:
\b \label{coneq}   \mathsf{h}(\Phi,\Pi,\nabla^\top_i\Phi)=\Pi \nabla^\top_0\Phi- \mathcal{L}(\Phi,\nabla^\top_\alpha\Phi)\,,\e
where $i=1,\cdots,4$. The explicit calculation of this function in the dS ambient space formalism for elementary fields is possible. Its physical meaning is unclear but at the null curvature limit it can be identified with the Hamiltonian density in Minkowski spacetime. 

From this fact and inspired by the Wheeler-DeWitt equation, we define the constraint equation of geometry quantum state as follows:
\b \label{cons2}  |\mathcal{G}\,;\mathcal{N}\rangle\in\mathcal{F}(\mathcal{K}^{(1)})  \Longrightarrow \mathsf{H} \vert \mathcal{G}\,;\mathcal{N} \rangle\equiv  \left( \mathsf{H}_f +\mathsf{H}_{int}\right)\vert \mathcal{G}\,;\mathcal{N} \rangle=0  \, ,\e
where $\mathsf{H}(\Phi,\Pi)\equiv \int  \ud\mu(x) \mathsf{h}(\Phi,\Pi,\nabla^\top_i\Phi)$. The first part is free fields theory which includes the dS massive gravity, the linear gravitational wave, and the mmc scalar field. The second part concerns the interaction of various fields. Using equation \eqref{teod} and \eqref{cons2}, we obtaine $\mathsf{H}\mathsf{U}\vert \mathcal{G}\,;\mathcal{N} \rangle=0=\mathsf{U}\mathsf{H}\vert \mathcal{G}\,;\mathcal{N} \rangle$. Therefore the simple form of $\mathsf{U}$, which satisfies the conditions \eqref{evolu}, is:
\b \label{evoph} \mathsf{U}(\mathcal{N},\mathcal{N}')\equiv e^{-\ii \int  \ud\mu(x) \mathsf{h}(\mathcal{N}-\mathcal{N}')}\,.\e
Although the physical meaning of $\mathsf{H}$ is unclear, it remains constant throughout the universe's evolution. By dividing it into geometrical and non-geometrical parts, $\mathsf{H}=\mathsf{H}_g +\mathsf{H}_{ng} \,$, we have a fluctuation between these two parts under the evolution of the universe, which neither is constant individually. It may be interpreted as an "energy" exchange between our universe's geometrical and non-geometrical parts. While the geometry quantum state evolves in Krein-Fock space, fluctuation of $\theta_{\alpha\beta}$ breaks the dS invariant. The explicit calculation of the equation \eqref{evoph} is out of the scope of this paper and will be discussed elsewhere. 

In summary, to construct the quantum geometry in this article, we have used four essential key ideas, briefly recalling them. 1) Utilising the ambient space formalism to attain an observer-independent perspective, which is essential for quantum geometry. 2)  Substituting Riemannian geometry with Weyl geometry to describe the spacetime geometry by the metric tensor and the mmc scalar field since the latter is also a geometrical field.  3) Replacing the Hilbert space with the Krein space to achieve a complete space and a covariant quantization. 4) The time parameter for quantum state evolution is replaced with the total number of quantum states to obtain an observer-independent formalism. 


\section{Conclusion}

In quantum dS geometry, the Hilbert space $\mathcal{H}$ is no longer a complete space. Instead, it is a subspace of a complete Krein space, $\mathcal{H}\subset \mathcal{K}$, in which the requirement for positive definiteness is abandoned. Replacing Hilbert space with Krein space is essential in our quantum geometry model. Krein space quantization (including quantum light cone fluctuation) permits us to construct a renormalizable QFT in curved space and quantum geometry.  Ambient space formalism permits us to formulate quantum geometry from an observer-independent point of view and to visualize the many-world interpretation. It should be noted that although the metric quantization breaks the dS invariant, the Krein-Fock space is a complete space for quantum geometry. The dS geometry quantum state is introduced as a superposition of the Krein-Fock space basis, and its evolution is parametrized in terms of the total number of quantum states. Using the idea of the Wheeler-DeWitt constraint equation in cosmology, the evolution equation of geometry quantum state can be written in terms of the Lagrangian density of interaction fields.

\vspace{0.5cm}
{\bf{Acknowledgments}}: The author wishes to express particular thanks to Jean Pierre Gazeau and Eric Huguet for their discussions. The author would like to thank Coll\`ege de France, Universit\'e Paris Cit\'e, and Laboratoire APC for their hospitality and financial support.



\begin{thebibliography}{99}

\bibitem{tagahu} M.V. Takook, J.P. Gazeau, E. Huguet, (2023), Universe 9(9): 379 (2023), \textit{Asymptotic states and S-matrix operator in de Sitter ambient space formalism},  [\href{http://export.arxiv.org/abs/2304.04756}{arXiv:2304.04756}].

\bibitem{taga} M.V. Takook, J.P. Gazeau, \href{https://doi.org/10.1016/j.nuclphysb.2022.115811}{Nucl. Phys. B {\bf 980}, 115811 (2022),} \textit{Quantum Yang-Mills theory in de Sitter ambient space formalism}, [\href{https://arxiv.org/abs/2112.02651v2}{arXiv:2112.02651v2}].

\bibitem{ta231} M.V. Takook, \href{https://iopscience.iop.org/article/10.1209/0295-5075/acb0f9}{Europhys. Lett. {\bf 1410}, 22003 (2023),} \textit{Axiomatic de Sitter quantum Yang-Mills theory with color confinement and mass gap}, [\href{https://arxiv.org/abs/2211.16060}{arXiv:2211.16060}].

\bibitem{gareta00} J.P. Gazeau, J. Renaud, M.V. Takook, Class. Quant. Grav. {\bf 17}, 1415 (2000), \textit{Gupta-Bleuler quantization for a minimally coupled scalar field in de Sitter space}, [\href{http://arxiv.org/abs/gr-qc/9904023}{gr-qc/9904023}].

\bibitem{ta22} M.V. Takook, Mod. Phys. Lett. A {\bf 37}, 2250059 (2022), {\it "Krein" regularization method}, [\href{https://arxiv.org/abs/2112.05390}{arXiv:2112.05390}].

\bibitem{morris} T.R. Morris, J. High Energ. Phys. {\bf 08}, 024 (2018), {\it Renormalization group properties of the conformal sector: towards perturbatively renormalizable quantum gravity}, [\href{https://arxiv.org/abs/1802.04281}{arXiv:1802.04281}].

\bibitem{anmo} I. Antoniadis, E. Mottola, Phys. Rev. D {\bf 45}, 2013 (1992), {\it Four-dimensional quantum gravity in the conformal sector}.

\bibitem{anmamo} I. Antoniadis, P.O. Mazur, E. Mottola, Phys. Rev. D {\bf 55}, 4770 (1997), {\it Physical States of the Quantum Conformal Factor}, [\href{https://arxiv.org/abs/hep-th/9509169}{hep-th/9509169}].

\bibitem{whe} J.T. Wheeler, Gen. Relativ. Gravit. {\bf 50}, 80  (2018), {\it Weyl geometry}, [\href{https://arxiv.org/abs/1801.03178}{arXiv:1801.03178}].

\bibitem{ta09} M.V. Takook, Iranian Physical Journal {\bf 3}, 1 (2009), {\it Linear gravity in de Sitter universe}, [\href{https://arxiv.org/abs/1710.06605}{arXiv:1710.06605}].

\bibitem{allen85} B. Allen, Phys. Rev. D {\bf 32}, 3136 (1985), {\it Vacuum states in de Sitter space}.

\bibitem{ta223} M.V. Takook, \href{https://www.sciencedirect.com/science/article/pii/S0550321322003170}{Nucl. Phys. B {\bf 984}, 115966 (2022)}, \textit{Scalar and vector gauges unification in de Sitter ambient space formalism}, [\href{https://arxiv.org/abs/2204.00314}{arXiv:2204.00314}].

\bibitem{chlopewi} V. Chandrasekaran, R. Longo, G. Penington, E. Witten, J. High Energ. Phys. {\bf 02}, 082 (2023), {\it An Algebra of Observables for de Sitter Space}, [\href{https://arxiv.org/abs/2206.10780}{arXiv:2206.10780}].

\bibitem{pewi}  G. Penington, E. Witten, (2023), {\it Algebras and States in JT Gravity}, [\href{http://export.arxiv.org/abs/2301.07257}{arXiv:2301.07257}]. 


\bibitem{ta20} M.V. Takook, Int. J. Theor. Phys. {\bf 59}, 2540 (2020), {\it Conceptual and Technical Challenges
of Quantum Gravity}, [\href{https://arxiv.org/abs/2208.08098}{ arXiv:2208.08098}].

\bibitem{gui} D. Guido, Contemp. Math. {\bf 534}, 97 (2011), \textit{Modular theory for the von Neumann algebras of Local Quantum Physics}, [\href{http://de.arxiv.org/abs/0812.1511v1}{arXiv:0812.1511v1}].

\bibitem{chsu} A. Chamseddine, W.D. van Suijlekom, A chapter of the book: \href{https://doi.org/10.1007/978-3-030-29597-4_1}{Advances in Noncommutative Geometry} (2019), {\it A survey of spectral models of gravity coupled to matter}, [\href{https://arxiv.org/abs/1904.12392}{arXiv:1904.12392}].

\bibitem{ta14} M.V. Takook, (2014), \textit{Quantum Field Theory in de Sitter Universe: Ambient Space Formalism}, [\href{http://arxiv.org/abs/1403.1204v3}{arXiv:1403.1204v3}].

\bibitem{khletu} D. Kharzeev, E. Levin, K. Tuchin, Phys. Rev. D {\bf 70}  054005 (2004), \textit{QCD in curved space-time: a conformal bag model}, [\href{https://arxiv.org/abs/hep-ph/0403152v1}{arXiv:0403152v1}].

\bibitem{bida} N. D. Birrell, P. C. W. Davies, Cambridge University Press (1982), {\it Quantum fields in curved space}.

\bibitem{gagata} T. Garidi, J. P. Gazeau and M. V. Takook, J. Math.
Phys. {\bf 44}, 3838 (2003), {\it Massive spin-2 field in de Sitter space}, [\href{http://arxiv.org/abs/hep-th/0302022}{arXiv:hep-th/0302022}]. 

\bibitem{gagarota} T. Garidi, J.P. Gazeau, S. Rouhani, M.V. Takook, J. Math. Phys. \textbf{49}, 032501 (2008), \textit{Massless vector field in de Sitter universe}, [\href{https://arxiv.org/abs/gr-qc/0608004}{arXiv:0608004v1}].

\bibitem{anilto} I. Antoniadis, J. Iliopoulos et T. N. Tomaras, Phys. Rev. Lett. {\bf 56}, 1319 (1986), {\it  Quantum instability of de Sitter space }.

\bibitem{tho} L.H. Thomas, \href{http://dx.doi.org/10.2307/1968990}{ Ann. of Math.} {\bf 42}, 113 (1941), {\it On unitary representations of the group of de Sitter space}.

\bibitem{dix} J. Dixmier, Bull. Soc. Math. France {\bf 89}, 9 (1961), {\it Repr\'esentation int\'egrables du group de de Sitter}.

\bibitem{tak} B. Takahashi, Bull. Soc. Math. France {\bf 91}, 289 (1963), {\it Sur les repr\'esentations unitaires des groupes de Lorentz g\'en\'eralis\'es}.

\bibitem{moy} P. Moylan, J. Math. Phys. {\bf 24}, 2706 (1983), {\it Unitary representation of the
(4+1)-de Sitter group on irreducible representation spaces of the Poincar\'e group}.

\bibitem{brmo96} J. Bros, U. Moschella, Rev. Math. Phys. {\bf 8}, 327 (1996), \textit{Two-point functions and
Quantum Field in the de Sitter Universe}, [\href{http://arxiv.org/abs/gr-qc/9511019}{gr-qc/9511019}].

\bibitem{ta3} M. V. Takook, Mod. Phys. Lett. A \textbf{16}, 1691 (2001), \emph{Covariant two point function for a minimally coupled scalar field in de Sitter spacetime}, [\href{https://arxiv.org/abs/gr-qc/0005020}{gr-qc/0005020}].

\bibitem{ta02} M.V. Takook, Int. J. Mod. Phys. E {\bf 11}, 509 (2002), {\it Negative Norm States in de Sitter Space and QFT without Renormalization Procedure}, [\href{https://arxiv.org/abs/gr-qc/0006019}{gr-qc/0006019}].

\bibitem{taen} M.V. Takook, Annals of Phys. {\bf 367}, 6 (2016), {\it Entropy of Quantum Fields in de Sitter Space-time}, [\href{http://arxiv.org/abs/1306.3575v2}{arXiv:1306.3575v2}].


\end{thebibliography}
\end{document}